\documentclass{emulateapj}
\usepackage[usenames,dvipsnames]{color}          

\shorttitle{Spectro-polarimetric simulations of solar limb}
\shortauthors{S.~Shelyag}

\begin{document}
\title{Spectro-polarimetric simulations of the solar limb: absorption-emission Fe\,{\sc i} $6301.5\mathrm{\AA}$ and $6302.5\mathrm{\AA}$ line profiles and torsional flows in the intergranular magnetic flux concentrations}

\author{S.~Shelyag}
\affil{School of Mathematical Sciences, Monash University, Victoria 3800, Australia}

\date{01.01.01/01.01.01}

\begin{abstract}
Using radiative magneto-hydrodynamic simulations of the magnetised solar photosphere and detailed spectro-polarimetric diagnostics with the
Fe\,{\sc i} $6301.5\mathrm{\AA}$ and $6302.5\mathrm{\AA}$ photospheric lines in the local thermodynamic equilibrium approximation, we 
model active solar granulation as if it was observed at the solar limb. We analyse general properties of the radiation across the solar limb, such 
as the continuum and the line core limb darkening and the granulation contrast. We demonstrate the presence of profiles with both emission 
and absorption features at the simulated solar limb, and pure emission profiles above the limb. These profiles are associated with the regions of
strong linear polarisation of the emergent radiation, indicating the influence of the intergranular magnetic fields on the line formation.
We analyse physical origins of the emission wings in the Stokes profiles at the limb, and demonstrate that these features are produced by 
localised heating and torsional motions in the intergranular magnetic flux concentrations. 
\end{abstract}


\section{Introduction}

The solar photosphere at the limb remains one of the largely unexplored solar regions due to the lack of bright and extended emissive features, such as spicules 
in the chromosphere, and sharp limb darkening, associated with low light conditions and high instrumental noise levels in the observations of this region. On 
the other hand, the solar limb has potential for the diagnostics of parallel to the solar surface photospheric flows and magnetic fields. Together 
with solar disk observations, it can provide new and more detailed information on the geometry of the flow and magnetic field structures in the solar photosphere, 
on the details of transition between the photosphere and the chromosphere, as well as on the origins of chromospheric features.

Groundbreaking spectro-polarimetric observation of the solar limb with Hinode SOT \citep{tsuneta2008} by \citet{lites2010} revealed a thin, sub-arcsecond layer 
of photospheric emission in the $6301.5\mathrm{\AA}$ and $6302.5\mathrm{\AA}$ lines of neutral iron. These lines, normally in absorption within the solar disk,
undergo a transition into pure emission above the limb. The transition between absorption and emission is characterised by the appearance of emissive wings in the
line profiles, their significant Doppler shifts, and linear (radially directed) light polarisation. 

Realistic numerical simulations of solar surface magneto-convection have long been successful in reproducing, explaining and even predicting various solar
photospheric \citep[see e.g.][and references therein]{stein2012}, and, recently, chromospheric phenomena \citep{leenaarts2012, pereira2013}. Centre-to-limb
variation of solar radiation has been extensively studied using simulations \citep{keller2004,koesterke2008,kitiashvili2014}. Ultimately, the success of numerical
modelling in understanding the physics of the solar photosphere can be demonstrated by an excellent agreement between the various numerical codes used to simulate 
the solar surface layers \citep{beeck2012}.

In this paper, we use numerical simulation of the solar photosphere with the MURaM code \citep{voegler1} and one-dimensional line profile synthesis with the NICOLE code 
\citep{socas2014} to perform the 
modelling of the spectro-polarimetric properties of the radiation at the solar limb with the $6301.5\mathrm{\AA}$ and $6302.5\mathrm{\AA}$ Fe\,{\sc i} lines in
the LTE approximation. We show the average properties of simulated radiation, such as centre-to-limb variation of the continuum intensity and the intensity in the
line core. We also show the connection between the line core intensity and the linear polarisation above the limb. At the limb, we demonstrate the presence of the transition layer
between pure absorption and pure emission, spectropolarimetrically similar to the observations by \citet{lites2010}. In the transition layer, the profiles with emission
wings are found. Using response functions to the temperature for the $6301.5\mathrm{\AA}$ line, we analyse the line formation process and provide a simple model for 
emission-absorption line profile formation at the solar limb. In our simulations, the emission wings in the profiles are produced by positive temperature gradients 
and torsional line-of-sight motions within the intergranular magnetic flux tubes. In contrast to the proposed role of non-LTE scattering \citep{lites2010} in formation 
of photospheric line profiles above the limb, we are able to reproduce the wing emission in the profiles in pure one-dimensional LTE approximation. We suggest that 
both non-LTE scattering and Doppler-shifted emission in the flux tubes are responsible for the complex Stokes-$I$ profile shapes at the solar limb.

The paper is organised as follows. In Section 2 we describe the simulation setup we used to model the solar limb radiation and polarization information.
Section 3 is divided into subsections 3.1 and 3.2, where we describe the general properties of the simulated limb, and analyse the physical origins of 
emission in the line profile wings at the limb, respectively. Section 4 concludes our findings.

\section{Simulation setup}

\begin{figure*}
\includegraphics{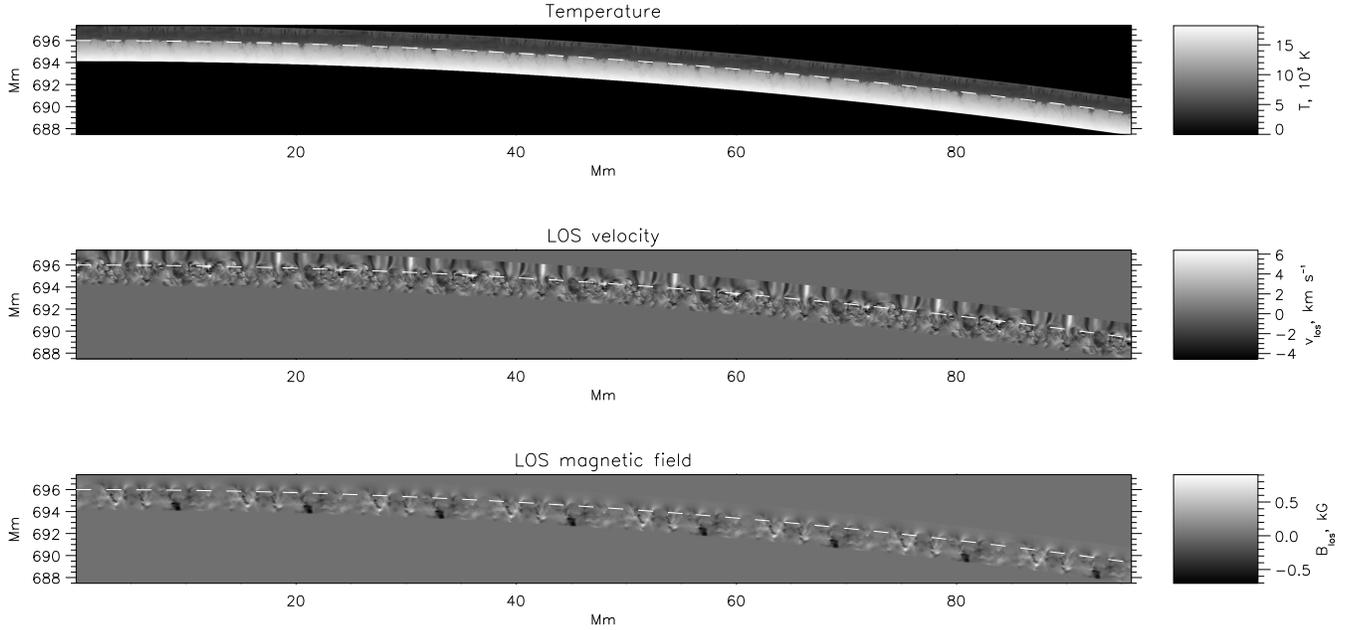}
\caption{Vertical cut through a half of the simulation domain representing the solar limb. The temperature, LOS velocity and 
LOS magnetic field strength are shown in the upper, middle and bottom panels, respectively. The vertical axis in the plots 
corresponds to the distance from the solar centre. The dashed line in the panels indicates approximate level of the solar surface.}
\label{fig1}
\end{figure*}

We use a single extended photospheric MHD model snapshot from a unipolar magnetic field simulation of a plage region produced by the MURaM code \citep{voegler1}. 
The code solves radiative MHD equations on a Cartesian grid.
The magneto-convection simulation is set through injecting a uniform magnetic field of $200~\mathrm{G}$ into a previously calculated non-magnetic
convection model. The resolution is $25~\mathrm{km}$ in the horizontal directions and $10~\mathrm{km}$
in the vertical direction. The simulation box has a size of $12 \times 12~\mathrm{Mm^2}$ and $3.2~\mathrm{Mm}$ in the horizontal
and vertical directions, respectively. The average continuum formation layer is located at about $2~\mathrm{Mm}$ above the bottom boundary. 
The bottom boundary of the domain is open for in- and outflows, and the top boundary is open 
for outflows, while the inflows are kept at constant temperature $6000~\mathrm{K}$ in order to mimick the chromospheric layers
of the Sun. The side boundaries are periodic. We use FreeEOS equation of state \citep{freeeos} to account for changes in the gas law due
to partial ionization. No Hall term or ambipolar diffusion is included in the simulation.

A solar limb region is constructed by adding together 16 copies of the chosen snapshot and locally changing the geometry of the snapshots to
cylindrical geometry with radius $R_\sun=695~\mathrm{Mm}$ in order to mimic the curvature of the solar surface in one dimension. The final size 
of the domain is about $192 \times 10~\mathrm{Mm}$ in the horizontal and vertical directions, respectively, which corresponds to $\pm8^\circ$ 
inclination between the surface and the line-of-sight (LOS), or $\mu=0-0.14$. The LOS components of the magnetic field and the plasma velocity 
are recalculated accordingly. After the domain is projected back to the Cartesian grid, it is resampled to a somewhat lower ($25~\mathrm{km}$) 
vertical resolution to reduce the computational cost, and smoothed along the lines-of-sight with
a boxcar function with the width of $100~\mathrm{km}$ to avoid numerical instabilities due to sharp gradients between granular and intergranular
solar magneto-convection features during radiative transport computations. With the optical ray length of about $5~\mathrm{Mm}$ the effect
of the smoothing is negligible.
Using the same model snapshot to cover the solar limb allows us to study the same features under the different inclination angles as well as at the solar limb.
A vertical cut through half of the constructed solar limb domain is shown in Fig.~\ref{fig1}. As is evident from the figure, in the constructed domain 
both the regions of the solar disk (with optically thick, hot and dense plasma) and the solar limb (optically thin) are present. The LOS components of the 
magnetic field and the velocity are predominantly horizontal in this simulation. Therefore, granular upflows and magnetised downflows do not influence 
the radiation. However, the photospheric part of the domain is covered with horizontally expanding magnetic fields and with the radially elongated 
line-of-sight velocity structures of alternating sign corresponding to Alfv\'enic horizontal torsional motions. These motions have been recently 
demonstrated to exist in the intergranular magnetic flux concentrations in simulations and analysed by \citet{shelyag2013,shelyag2014}.

In order to calculate Stokes profiles for $6301.51\mathrm{\AA}$ and $6302.49\mathrm{\AA}$ Fe\,{\sc i} absorption lines, the horizontal one-dimensional 
columns from the solar limb model were used as an input for spectro-polarimetric line profile synthesis code NICOLE \citep{socas2000, socas2011,socas2014}. 
No multi-dimensional scattering or non-LTE effects were included in the computation.  The excitation potentials $V=3.65~\mathrm{eV}$ and 
$\log{gf}=-0.59$, and $V=3.686~\mathrm{eV}$ and $\log{gf}=-1.13$ were used for $6301.51\mathrm{\AA}$ and $6302.49\mathrm{\AA}$ Fe\,{\sc i} lines 
synthesis, respectively. The full Stokes vectors for the lines were calculated within $6301.2\mathrm{\AA}-630\mathbf{1}.8\mathrm{\AA}$, and 
$6302.2\mathrm{\AA}-6302.8\mathrm{\AA}$ wavelength ranges with $3~\mathrm{m\AA}$ resolution.

\section{Results}
\subsection{General properties}

\begin{figure}
\plotone{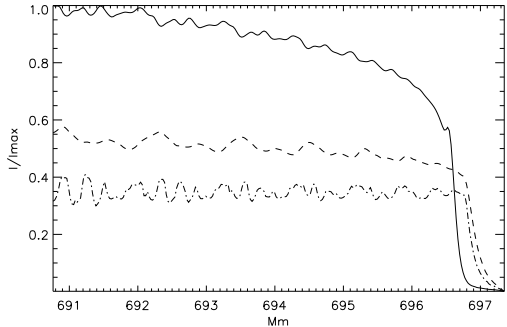}
\caption{Limb darkening across the solar limb. $6300\mathrm{\AA}$ continuum intensity is shown with the solid line, the dashed line corresponds to 
the intensity at the wavelength of the line core $6302.49\mathrm{\AA}$, and the dash-dotted line corresponds to the minimum intensity of the Fe\,{\sc i} 
$6302.49\mathrm{\AA}$ line profile. The horizontal axis is the distance from the solar centre.}
\label{fig2}
\end{figure}

\begin{figure}
\plotone{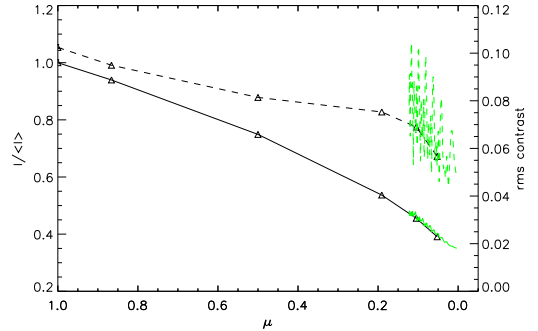}
\caption{$6300\mathrm{\AA}$ continuum limb darkening across the solar disk. Solid curve - simulated average normalized intensity, dashed curve - {\it rms} contrast 
of granulation. The solar limb model ($\mu=0.15-0$) is shown in green. Triangles show the inclination angles used to calculate the radiation parameters,
$\mu=1,~\sqrt{3}/2,~0.5,~0.2,~0.1,~0.05$.}
\label{fig3}
\end{figure}

\begin{figure*}
\plotone{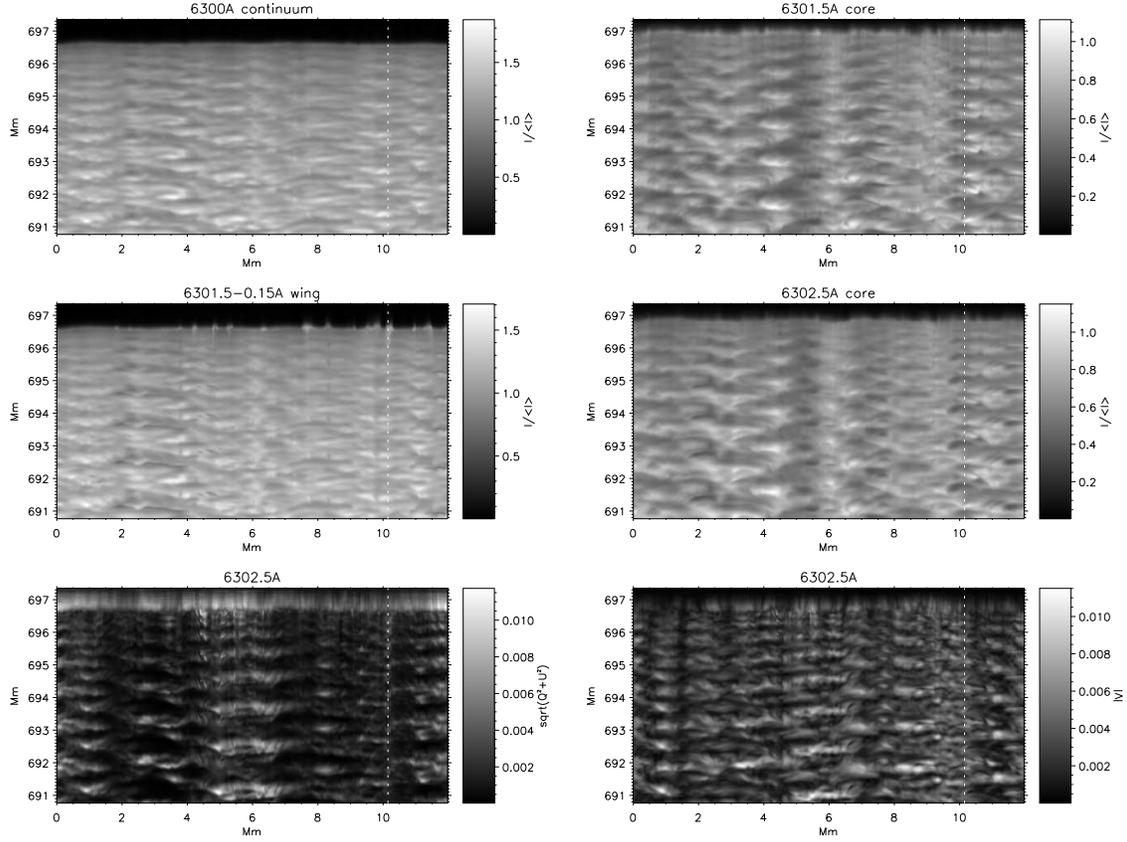}
\caption{Continuum (top left), $6301.5\mathrm{\AA}$ line core (top right), $6301.5-0.15\mathrm{\AA}$ line wing (middle left), $6302.5\mathrm{\AA}$ (middle right) 
line core normalised intensity images from the simulations, and $\int\sqrt{Q^2+U^2}d\lambda$ (bottom left) and $\int|V|d\lambda$ (bottom right) simulated images.
Small-scale local brightenings are visible in $6301.5\mathrm{\AA}$ line wing, and, to some extent, in the line cores. Integrated Stokes parameters show presence of 
both circular and linear polarisation. The linear polarisation is significantly more pronounced above the simulated limb compared to the circularly polarized light in the 
same region.}
\label{fig4}
\end{figure*}

\begin{figure*}
\plotone{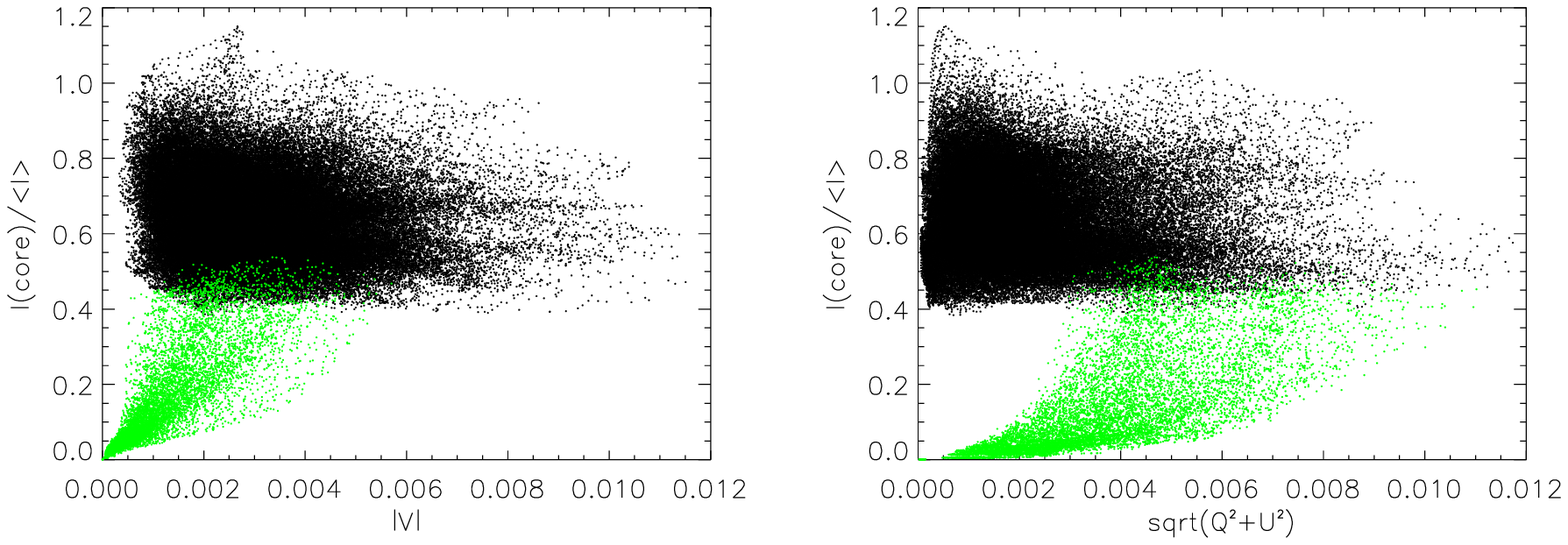}
\caption{Scatter plots of intensity in the line core vs the total Stokes-$V$ area (left panel) and of intensity in the line core vs the total Stokes-$\sqrt{Q^2+U^2}$ area
(right panel) for Fe\,{\sc i} $6302.5\mathrm{\AA}$ line. The green dots correspond to the $r > 696.8~\mathrm{Mm}$ in Fig.~\ref{fig4}.}
\label{fig5}
\end{figure*}

The average radiation intensity structure across the simulated limb is shown in Fig.~\ref{fig2}. The solid curve represents the normalised continuum intensity variation.
A region with sharp intensity decrease from $0.5$ to nearly zero at $696.6~\mathrm{Mm}$ has thickness of about $200~\mathrm{km}$. 
At the rest wavelength of the Fe\,{\sc i} line core, $\lambda=6302.49\mathrm{\AA}$, the average intensity (dashed curve) is about $0.5$ of the continuum 
intensity at the solar disk. However, due to the increased opacity in the line core, the simulated solar limb is higher in the solar atmosphere by about 
$300~\mathrm{km}$ and spreads over a somewhat larger height range of about $300~\mathrm{km}$. Similar behaviour is demonstrated by the minimum
intensity in the line profile (shown as dash-dotted curve). It should be also noted that the averaging has been done over a single computational snapshot,
which does not account for the oscillations and time dependence present in the simulations. Therefore signatures of intensity variability due to granulation
are also present.

Average continuum intensites across the full solar disk were also computed. Fig.~\ref{fig3} shows the average $6300\mathrm{\AA}$ normalised intensities
calculated for the same model snapshot at different observational angles at the solar disk $\mu=1,~\sqrt{3}/2,~0.5,~0.2,~0.1,~0.05$ using the domain inclination 
method \citep[see e.g.][black solid line]{shelyag2014}, the 
average intensities from the limb simulation ($\mu=0.14-0$; green solid line), the granulation contrast (black dashed line) and the granulation contrast from 
the limb simulation (green dashed line). As the plot shows, the limb darkening dependences on $\mu$ fit exactly in the overlapping region between
$\mu=0.05-0.1$. The {\it rms} contrast dependence on $\mu$ for the limb simulation shows large scatter due to significantly smaller statistics. However,
the scatter is found to be close around the average contrast values for the solar disk computation.

A continuum intensity image (Fig.~\ref{fig4}, top-left panel) shows low-contrast solar limb granulation and a corrugated solar limb surface. The limb height variability 
due to granulation is small, of the order of $100-200~\mathrm{km}$, as the (deeper) intergranular lanes are completely hidden by the granulation.
Bright granular walls (faculae) are visible \citep{keller2004}. There are no noticeable continuum intensity features above the solar limb. The $6301.5\mathrm{\AA}$ line core intensity 
image (top-right panel in Fig.~\ref{fig4}) and, more pronounced, the $6301.5-0.15\mathrm{\AA}$ line wing intensity image (middle-left panel), however, show 
a large number of small-scale vertically extending features with somewhat enhanced line core and strongly enhanced line wing intensities above the average solar limb radius
at this wavelength, $R>697~\mathrm{Mm}$. The $6302.5\mathrm\AA$ line core intensity (middle-right panel) shows similar, although less prominent features. Reversed granulation 
is also visible \citep{cheung2007}.

\begin{figure}
\plotone{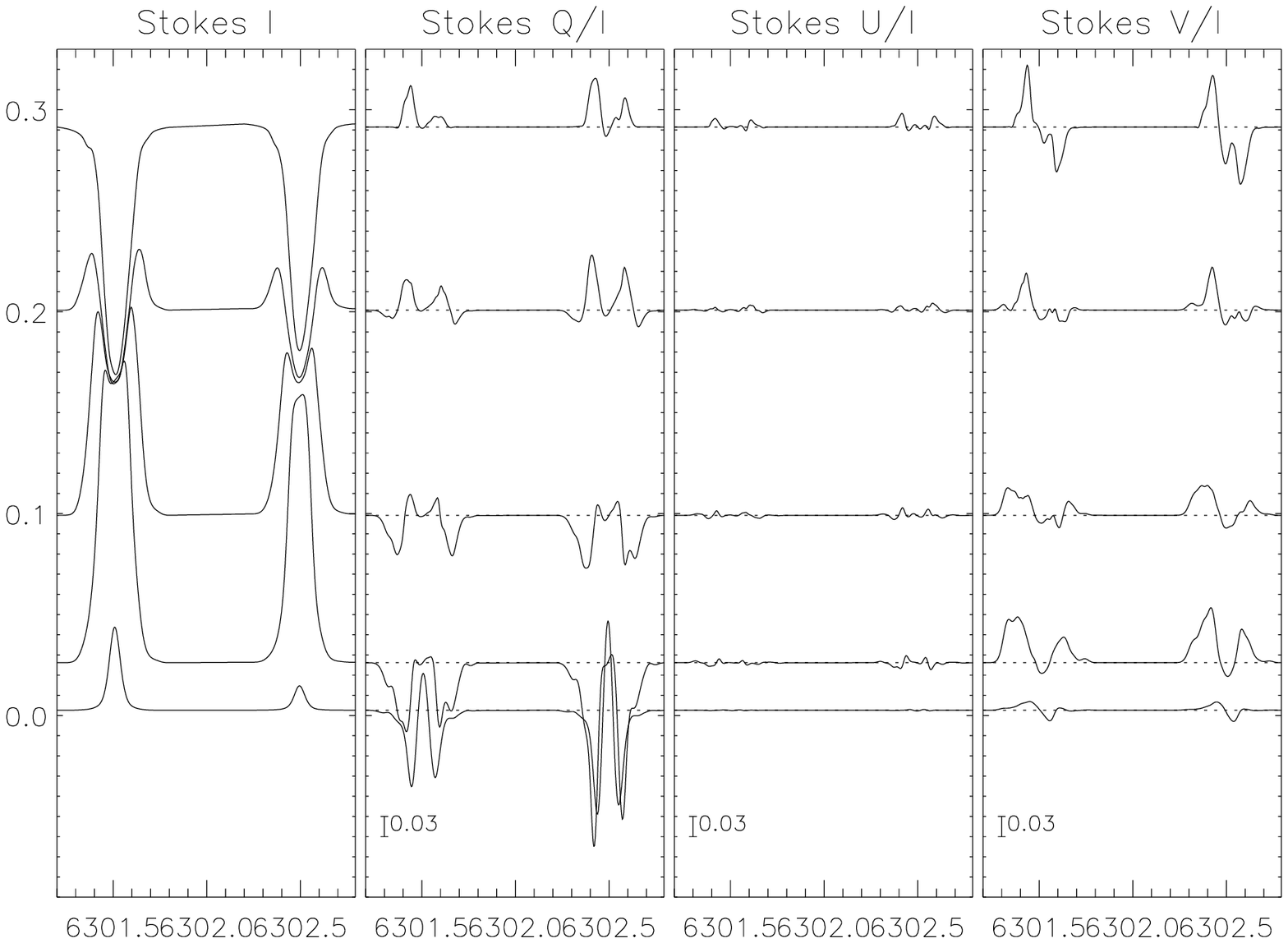}
\caption{Horizontally averaged Stokes-$I$, $Q/I$, $U/I$, $V/I$ profiles for the radii $R=695.875,~696.6,~696.675, ~696.775,~697.2~\mathrm{Mm}$ (from the top to the bottom) 
in the model. The dotted lines show zero levels for $Q/I$, $U/I$ and $V/I$ profiles, which are normalised so one tick interval in the corrsponding plots is equal to $0.03$.}
\label{fig6}
\end{figure}

As can be seen from the bottom panel of Fig.~\ref{fig1}, the LOS magnetic field mainly corresponds to the horizontal magnetic field of the intergranular
magnetic field concentrations, expanding in the solar atmosphere. Therefore, along the line of sight, both polarities of the magnetic field, as well as a strong
component of the magnetic field perpendicular to the line of sight, are present. Together with torsional oscillatory flows in the magnetic flux tubes, this  leads to
generation of strong and asymmetric, multi-lobed profiles in all polarization components, as it is demonstrated in the next subsection.

The line parameters, namely the linear ($\int\sqrt{Q^2+U^2}d\lambda$) and circular ($\int|V|d\lambda$) polarisation, are shown in the bottom-left and bottom-right panels 
of Fig.~\ref{fig4}, respectively. The images show the presence of strong Stokes signals both at the solar limb and in the simulated solar disk region. As expected due to the 
weakness of the LOS magnetic field and strong magnetic field component perpendicular to the line-of-sight above the limb, the signal in Stokes-$V$ becomes weaker, and 
the signal in $\sqrt{Q^2+U^2}$ becomes stronger above the solar limb region. 

Scatter plots of the intensity in the line core versus total Stokes-$V$ and $\sqrt{Q^2+U^2}$ areas are shown in the left and right panels of Fig.~\ref{fig5}, respectively.
Both figures clearly show two separate populations of points: above the solar limb (marked by green dots) and at the solar disk (black dots). While the solar disk
population does not show a dependence of the Fe\,{\sc i} $6302.5\mathrm{\AA}$ line core intensity on the magnetic field, the solar limb population demonstrates 
such dependence: the larger the linear and/or circular polarisation, the higher the line core intensity is.
 
The horizontally averaged Stokes-$I$, $Q/I$, $U/I$, and $V/I$ profiles for $6301.5\mathrm{\AA}$ and $6302.5\mathrm{\AA}$ lines across the solar limb are shown in Fig.~\ref{fig6} for 
the radii $R=695.875,~696.6,~696.675, ~696.775,~697.2~\mathrm{Mm}$. The pure absorption profiles are obtained for the position within the simulated solar disk 
at $R=695.875~\mathrm{Mm}$. Stokes-$Q$ and $V$ show the presence of strong linear (vertical) and line-of-slight polarisation signals. At the limb, $R=696.6~\mathrm{Mm}$, 
Stokes-$I$ profiles show an absorption core, while the profile wings show emission. 

Stokes-$V$ profiles have lower amplitude compared to the disk due to the inclination of magnetic field with respect to the LOS, and the multi-lobed structure of the profiles 
becomes more pronounced due to presence of both positive and negative polarities of the magnetic field and velocity variations along the LOS and within the horizontal averaging element.

Above the limb, Stokes-$I$ become gradually more emissive in the optically thin part of the simulated atmosphere. 
In agreement with the observations \citep{lites2010}, the $6302.5\mathrm{\AA}$ line profile becomes entirely emissive closer to the limb than the $6301.5\mathrm{\AA}$ profile. 
At about $300~\mathrm{km}$ above the simulated limb both lines are in the emission regime. In the topmost layers of the simulation domain, both lines become fully emissive with no core absorption. The line intensity ratio of 3.4, which is close to the ratio of the values of the oscillator strengths used for the lines $\sim3.47$ (see Section 1) indicates optically thin emission regime of the lines. Stokes-$Q$ profiles demonstrate strong linear polarisation above the limb, while Stokes-$V$ becomes somewhat smaller, complex and multi-lobed, indicating that the LOS velocity and magnetic field experience significant changes in the region of the line formation. Stokes-$U$ does not show any significant horizontal polarisation due to the geometry of the magnetic field in the simulation.

\citet{lites2010} suggested that the Fe\,{\sc i} $6301.5\mathrm{\AA}$ and $6302.5\mathrm{\AA}$ line profile reversals to the emission regime in the wings in proximity to the solar limb 
are caused by non-LTE excitation of iron atoms by scattered light coming from the photosphere. Our simulation, however, shows the presence of such horizontally average 
profiles, computed using the pure one-dimensional LTE approximation. To determine the physical mechanism of the wing emission, we first performed the computation of the Fe\,{\sc i} line 
profiles in the horizontally averaged model with zero LOS velocity. This simple numerical experiment showed the absence of emission wings in the computed
line profiles and LTE-type transition of the lines from the absorption to emission regime across the simulated solar limb. Therefore, in the simulation we present here, the wing 
emission in the horizontally-averaged profiles is created by the superposition of the local, possibly asymmetric emission-absorption profiles produced by strong LOS 
(tangential to the solar surface) velocity and temperature gradients at the corrugated surface of the solar limb. In the next subsection, we analyse the details of formation of 
these profiles in our simulations.

\subsection{Absorption-emission profiles at the solar limb}

\begin{figure*}
\center
\includegraphics{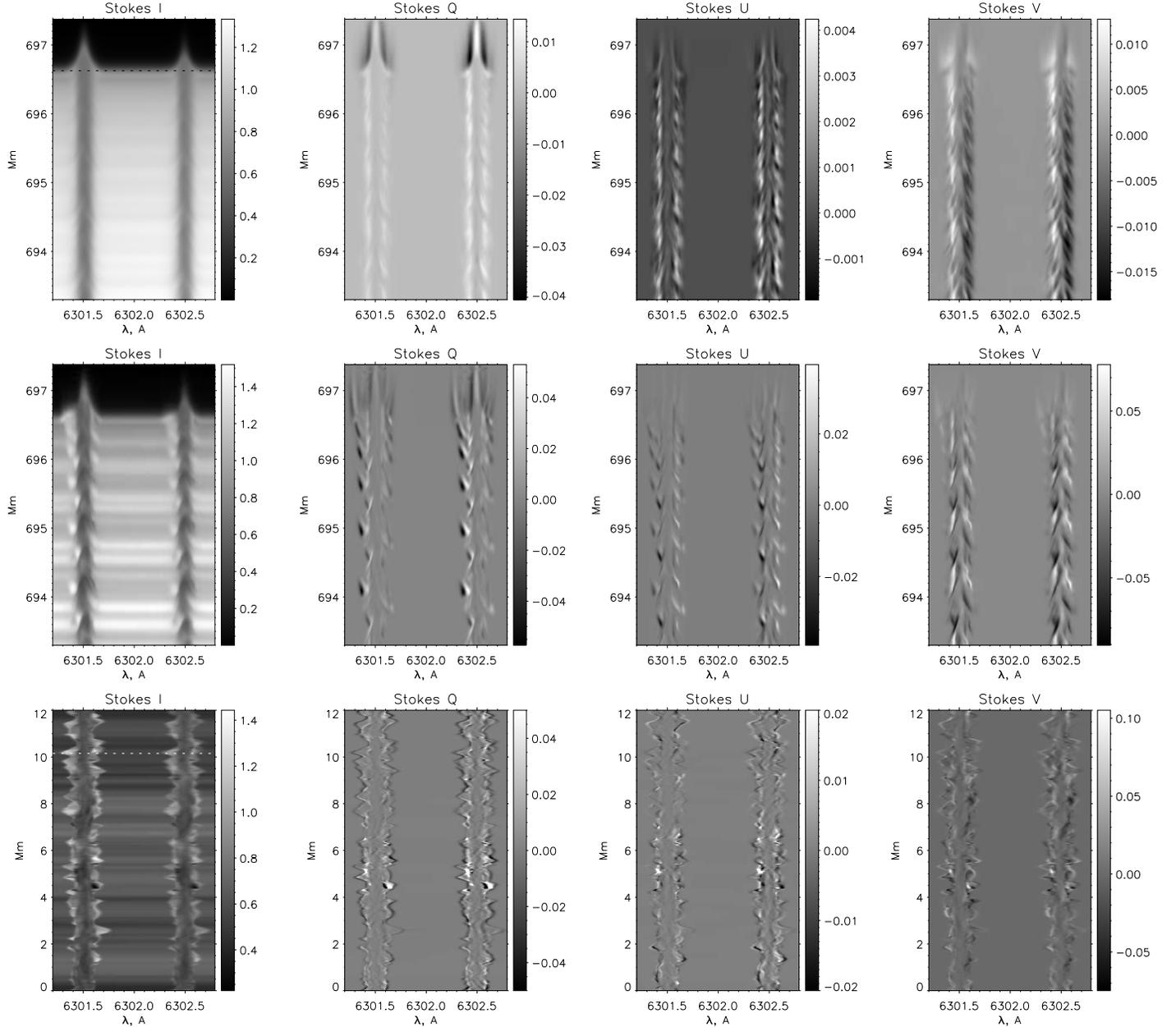}
\caption{Simulated $I$, $Q/I$, $U/I$, $V/I$ slit images for the horizontally averaged profiles (top row), for a position across the simulated solar limb marked by the dotted line 
in Fig.~\ref{fig4} (middle row), and for a slit positioned horizontally at the limb (bottom row, marked in the top row Stokes-$I$ image with the dashed line). The position of the vertical slit 
shown in the middle row is marked by the horizontal dashed line in the bottom row.}
\label{fig7}
\end{figure*}

Figure~\ref{fig7} shows simulated $I$, $Q/I$, $U/I$ and $V/I$ slit images for the horizontally-averaged simulated profiles (top row), for a slit positioned vertically across the 
solar limb at the position marked by the vertical dotted line in Fig.~\ref{fig4} (middle row), and for the slit positioned horizontally at the solar limb, as marked in the top-left (Stokes-$I$)
panel of the figure (bottom row). 

The horizontally-averaged Stokes-$I$ profiles show absorption within the simulated solar disk, the emission profiles far above the solar limb, and 
the transition from pure absorption to pure emission, which is characterised by the presence of the profiles with the absorption core and the raised, emissive wings. Stokes-$Q/I$
profiles within the disk are positive predominantly because the magneto-convection model used in this simulation has a unipolar vertical magnetic field. Above the limb, Stokes-$Q/I$ 
change their signs. Horizontally-averaged Stokes-$U/I$ profiles generally have multi-lobed shapes and lower amplitudes, as has already been shown, due to the absence of a preferred 
direction for the magnetic fields parallel to the limb in the simulation. Stokes-$V/I$ profiles, while also being complex and multi-lobed, show some preference towards positive
blue lobes at the disk, which is consistent with the geometry of magnetic field in opening magnetic flux tubes of positive polarity.

For the vertically-aligned slit, the Stokes-$I$ slit image shows strong continuum variability on the disk and emission cores above the limb for both the lines. However, the most pronounced 
feature in the image is the presence of bright features in the blue wings\footnote{Note that the presence of only the blue wing emission here is purely a selection effect introduced by 
choosing a position with a bright feature in the blue wing of the profile. Similarly frequent features in the red wing are also found in the simulations, as well as the features with two
emissive lobes of lower amplitude.} of both $6301.5$ and $6302.5\mathrm{\AA}$ Fe\,{\sc i} lines 
with the intensities higher and spatially unrelated to the continuum intensity enhancements. At the same locations as the bright features in the Stokes-$I$ wings, Stokes-$Q/I$ and $U/I$ show 
strong polarisation signals and rapidly change sign within the same intensity feature. Notably, Stokes-$Q/I$ and $U/I$ are, in general, of similarly large amplitudes ($\pm0.4$), 
indicating the presence of strong perpendicular to the LOS magnetic fields. Above the limb, Stokes-$Q/I$ show a multi-lobed structure, which does not vary with height 
significantly, while Stokes-$U/I$ vanishes. Stokes-$V/I$ signals within the disk show multiple lobes of both polarities too, indicating the presence of the gradients of LOS velocity 
and magnetic fields \citep[see e.g.][]{shelyag07}. It should be noted that no averaging or image degradation has been applied for the shown vertically-aligned slit, 
therefore the only source of strongly asymmetric multi-lobed Stokes-$V$ profiles is the variation of the physical parameters along the LOS.

Above the limb, Stokes-$V/I$ vanishes in agreement with the bottom-right panel of Fig.~\ref{fig4}.

The Stokes-$I$, $Q/I$, $U/I$ and $V/I$ images for the horizontally-positioned slit at the solar limb are shown in the bottom row of Fig.~\ref{fig7}. The Stokes-$I$ image shows the presence
of profiles with emission wings almost everywhere along the slit for both lines. It should be noted that most of the profiles are strongly asymmetric, with emission either in the
blue or in the red lobe. The presence of the emission lobes shows some correlation with the continuum intensity variation. The emission lobes are strongly shifted with the corresponding
LOS velocities of $4-8~\mathrm{km~s^{-1}}$. There is a weak visual correlation between Stokes-$Q/I$, $U/I$ amplitudes and the emission lobes, while Stokes-$Q/I$ has 
expectedly larger amplitude than Stokes-$U/I$. Nevertheless, the Stokes-$Q/I$ and $U/I$ signal patterns coincide with the emission lobes of the profiles, while there are no strong
signals in the line absorption cores. A similar dependence is found for the Stokes-$V/I$ signal.

\begin{figure*}
\center
\includegraphics{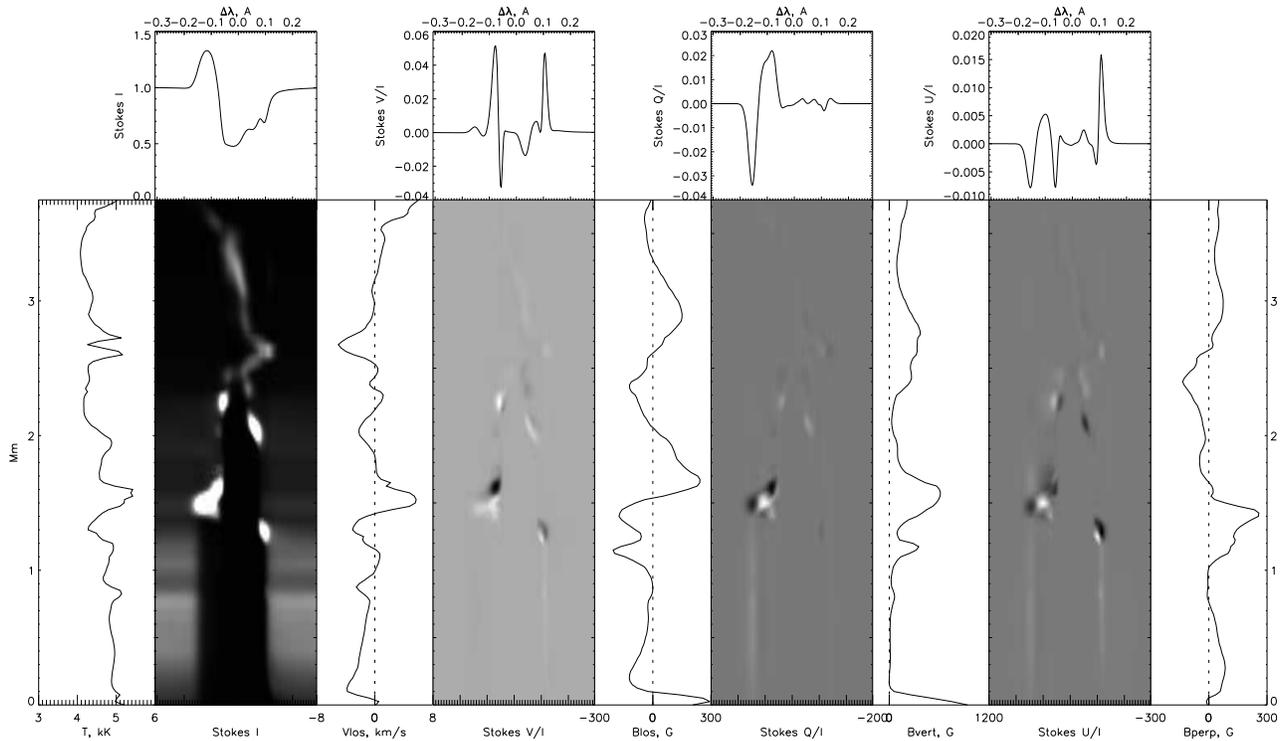}
\caption{Simulated $I$, $Q/I$, $U/I$, $V/I$ profiles (top row), their response functions to the temperature perturbation (grayscale images) and the structure of the 
 corresponding 1D atmosphere along the LOS. Dashed vertical lines mark zero values for $v_{los}$, $B_{los}$, $B_{|}$ and $B_{\perp}$.}
\label{fig8}
\end{figure*}

The process of formation of the profiles which show both emission and absorption is analysed below. An arbitrary position within the simulated solar disk, but close to the limb
was selected, where the $6301\mathrm{\AA}$ line profile shows both absorption in the line core and emission in the blue wing. Response functions (RFs) of all Stokes parameters 
to a small temperature perturbation ($\Delta T=100~\mathrm{K}$)  in the corresponding one-dimensional atmosphere were computed. Fig.~\ref{fig8} shows the calculated RFs
(grayscale images) together with the corresponding line profiles (plots in the top row), and the photospheric parameters (vertical plots). A clear correspondence is seen between the
emission part of the profile at $\Delta\lambda=-0.1\mathrm{\AA}$, and a sharp temperature rise at about $1.5~\mathrm{Mm}$ distance along the LOS. At the same location,
a strong plasma flow of about $4-5~\mathrm{km~s^{-1}}$ towards the observer is found, which shifts the emission wavelength towards the blue side of the spectrum. The
LOS component of the magnetic field $B_{los}$ changes the direction and amplitude from $-200~\mathrm{G}$ to $250~\mathrm{G}$, leading to the formation of a strong
and asymmetric Stokes-$V/I$ signal at the same wavelength \citep[see e.g.][]{shelyag07}. The vertical, perpendicular to the LOS component of the magnetic field
$B_|$ is almost always positive with the minimum magnetic field of a few Gauss. In the emission region it increases up to about $600~\mathrm{G}$, indicating that the  
line-of-sight crosses an intergranular magnetic field concentration there. The Stokes-$U/I$ RF has a maximum in this region, and the Stokes-$U/I$ profile shows two lobes of
large amplitude in the blue. The horizontal, perpendicular to the LOS magnetic field component $B_\perp$ has expectedly smaller amplitude and also shows an increase 
in the emission region up to $250~\mathrm{G}$.

\begin{figure}
\center
\includegraphics[width=0.5\textwidth]{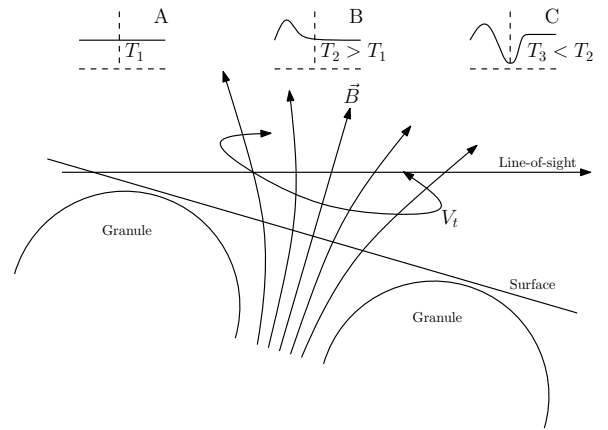}
\caption{Emission-absorption line profile formation in the non-uniform solar photosphere at the limb. The line-of-sight close to the solar limb crosses three different regions: (A) a non-magnetic
granular region, where the continuum radiation is formed, (B) the intergranular magnetic field concentration, where the emission part of the profile is formed, and which is characterised 
by radiative heating, positive temperature gradient  ($T_2>T_1$), and torsional plasma motions ($V_t$), and (C) a non-magnetic region over the granule with a negative temperature 
gradient ($T_3<T_2$), where the absorption part of the profile is formed.}
\label{fig9}
\end{figure}

A summary of the process of photospheric emission-absorption Fe\,{\sc i} line profile formation is provided in Fig.~\ref{fig9}. The line-of-sight crosses three distinct regions. The
first region (A) is above the granule, where both the LOS components of the velocity and the magnetic field are small (from 0 to about $1.0~\mathrm{Mm}$ along the
LOS in Fig.~\ref{fig8}. In this region, the continuum radiation is formed. Then, the line-of-sight crosses the intergranular magnetic field concentration region (B).
Strong torsional oscillatory motions with speeds of a few $\mathrm{km~s^{-1}}$ within the magnetic concentration \citep{shelyag2013,shelyag2014} Doppler-shift the line formation
wavelength in accordance to the direction of motion in this region (from 1.0 to $2.0~\mathrm{Mm}$ in Fig.~\ref{fig8}). At the same time, radiative heat exchange with the hot granular wall 
\citep[see e.g.][]{spruit1976,voegler2004} and resistive heating \citep{moll2012,shelyag2014} cause the localised temperature increase within the magnetic flux tube 
and emission. Finally, the line-of-sight leaves the magnetic field concentration and enters the rarified region above the next granule (C), where the LOS velocities
are generally smaller and temperature decreases (up to the chromospheric layers, to which the photospheric lines under consideration are not sensitive), leading to the formation
of an absorption line core.

\section{Discussion}

In this paper, we presented a novel spectro-polarimetric simulation of photospheric radiation at the solar limb. We used a single snapshot from magneto-convection simulations
of a plage region to construct a cylindrical solar limb-like region and performed detailed radiative diagnostics computation on this domain with the $6301.5\mathrm{\AA}$ and 
$6302.5\mathrm{\AA}$ pair of photospheric lines of neutral iron in the LTE approximation. We computed the average radiation parameters at the transition between the solar disk and limb, 
such as limb darkening, $rms$ granulation contrast and average Stokes profiles. 

The average Stokes profiles showed the emission above the limb, the presence of emissive line wings, absorption core and significant polarisation 
signal at the limb, similar to the observations \citep{lites2010}, where it was suggested that the emissive wings are the result of radiation scattering and non-LTE excitation in the solar 
atmosphere. Our results, however, suggest that the radiation scattering at the limb is not the only mechanism causing the emission in the line wings. It is demonstrated that,
in the LTE approximation, the emission lobes in Stokes-$I$ profiles at the limb can be formed due to the presence of the torsional bidirectional flows along the line-of-sight 
and the local heating in the 
intergranular magnetic field concentrations. When the line-of-sight, along which the radiation is formed, crosses the intergranular magnetic concentration region, it experiences
a temperature rise, leading to emission, and a Doppler shift to the blue (or red) side of the spectrum. After that, the line-of-sight crosses an absorption region with a smaller
LOS velocity and a temperature decrease, which leads to the formation of the absorption core. As there is no preferred direction for the torsional flows in magnetic field
concentrations, both blue- and red-wing emission is observed in the simulations. The profiles with two emissive lobes are also found in the regions where the line-of-sight
crosses the layers of opposite rotation in the magnetic flux tubes \citep{shelyag2011b}. When averaged spatially, these profiles lead to the appearance of the Stokes-$I$ profiles
with emissive lobes and absorption core. The polarisation signals at the limb show the presence of vertically (radially) oriented magnetic fields, which is expected since a plage 
model with $200~\mathrm{G}$ vertical magnetic field was used for the computation. 

Despite the detailed characteristics of torsional motions in magnetic field concentrations still need to be better understood \citep[see e.g.][]{wedemeyer2014}, 
there is little doubt of their presence in the magnetic field concentrations. Furthermore, non-magnetic or weakly magnetised photospheric models also show torsional motions 
in intergranular lanes \citep{shelyag2011a,moll2011,kitiashvili2012,kitiashvili2013}. While in this case of zero or weak photospheric magnetic field the magnetic tension 
is small and does not prevent tornado-like appearance of the vortex, the purely hydrodynamic vortex flows in the intergranular lanes would lead to the presence of torsional 
motions consequently leading to the similar mechanism of emission-absorption line profile formation at the solar limb.

A natural point of concern is the use of LTE approximation in the presented study. While there is no doubt the non-LTE effects are important especially at the limb, the primary aim
of the study was to demonstrate the LTE effects of the limb radiation formation. Furthermore, the neutral iron lines used for spectropolarimetric diagnostics in this paper are
formed low in the photosphere irrespectively of the position at the solar disk, where the LTE assumption is still reasonable, at least for the intergranular lanes \citep{shchukina2001}. 
The basic process of emission-absorption line profiles formation at the limb, which is described in the paper, relies on the presence of torsional motions and heating within the intergranular
magnetic flux tubes.

The structures described in the paper are rather small-scale, with linear dimensions of about $100~\mathrm{km}$. Taking into account low light and high noise levels, inevitably
associated with limb observations, it would be difficult to provide detailed and direct observational confirmation of the effect of torsional flows in magnetic field concentrations on
the photospheric line formation at the solar limb. Nevertheless, future, large-scale instruments, such as DKIST, will be able to observe these small features. Finally, further study
is needed and planned to analyse their lifetimes and links to chromospheric fine structure, bidirectional flows and oscillations \citep{sekse2013}.

\acknowledgments
\section{Acknowledgement}
The author thanks the anonymous referee for his valuable comments.
This research was undertaken with the assistance of resources provided at the NCI National Facility systems at the Australian National University through the 
National Computational Merit Allocation Scheme supported by the Australian Government, and at the Multi-modal Australian ScienceS Imaging and Visualisation 
Environment (MASSIVE) (www.massive.org.au). The author also thanks Centre for Astrophysics \& Supercomputing of Swinburne University of Technology (Australia) 
for the computational resources provided. Dr Shelyag is the recipient of an Australian Research Council's Future Fellowship (project number FT120100057).

\end{document}